\documentclass[showpacs,showkeys,aps]{revtex4}
\usepackage{amssymb}
\usepackage{graphicx}
\usepackage{dcolumn}
\usepackage{bm}

\input{tcilatex}

\begin{document}

\title{The classical capacity for the quantum Markov channel of continuous
variables}
\author{Tao Qin$^{1}$, Meisheng Zhao$^{1}$ and Yongde Zhang$^{2,1}$}
\affiliation{$^{1}$Department of Modern Physics, University of Science and Technology of
China, Hefei 230026, People's Republic of China\\
$^{2}$CCAST (World Laboratory), P.O. Box 8730, Beijing 100080, People's
Republic of China}
\date{November 25, 2005}

\begin{abstract}
Quantum communications using continuous variables are quite mature
experimental techniques and the relevant theories have been extensively
investigated with various methods. In this paper, we study the continuous
variable quantum channels from a different angle, i.e., by exploring master
equations. And we finally give explicitly the capacity of the channel we are
studying. By the end of this paper, we derive the criterion for the optimal
capacities of the Gaussian channel versus its fidelity.
\end{abstract}

\maketitle

% ----------------------------------------------------------------

\section{Introduction}

The evaluation of information capacities of quantum channels is one of the
most challenging and intractable questions of quantum information theory
\cite{chuang}. Previous research primarily focuses on discrete input
alphabet. However, quantum information transmission with continuous alphabet
is an interesting alternative to the classical discrete alphabet based
approach \cite{pati,cover}. Many efforts have been devoted to characterize
continuous alphabet quantum channels \cite{bennett}. It is also important to
investigate the capacity of the continuous-variable quantum channels acting
on a bosonic field. Such issue has been addressed recently \cite%
{holevo1,giovannetti,macchiavello}. All these intriguing works mentioned
above have brought significant progress in the studies of
continuous-variable quantum channels.

One rationale is that any quantum operation is a completely positive
trace-preserving map(CPT) and therefore it can be considered as a quantum
channel. On the other hand, when the physical system of interest interacts
with the environment, irreversible decoherence can occur, causing the pure
states to become mixed states \cite{zurek,joos}. This process depicts the
influence of noise over quantum states, which can be envisioned as
transmission of information under noisy circumstances. Sonja \textit{et al.}
have investigated these kinds of noisy quantum channels for qubits, which
they call the squeezed vacuum channel, by making use of the master equation
\cite{sonja}. These studies provide a unique angle to address this issue.
However, their work is restricted to noisy quantum channels acting on
finite-dimensional Hilbert space, while the studies on continuous variable
system by using the master equation are obscure.

In this paper, we investigate the master equation of the decaying monomodal
electromagnetic field interacting with the thermal reservoir. We consider
the evolution of the density operator as a kind of information transmission
process undergoing a noisy quantum Markov channel. The channel we study is
Markovian, so memory is not an issue here \cite{markov}.

In the article we will give explicitly the capacity of the Markov channel.
The material is organized as follows. We begin this paper by introducing the
quantum Markov channel and its general properties in the first part of Sec.
II. The explicit solution to the master equation is discussed in the second
part of Sec. II. In the third part of Sec. II, we give detailed calculation
of the capacity of the quantum Markov channel. In Sec. III, some discussions
are made. The capacity result is analyzed. In Sec. IV, we will discuss the
fidelity of the channel transmission and derive for the particular input
signal such that the channel capacity and channel fidelity are mutually
optimal. And in Sec. V, we will arrive at the final conclusions.

\section{Quantum Markov channels}

\subsection{General properties of the Markov channel}

Generally speaking, any quantum physical operation that reflects the time
evolution of a quantum state can be regarded as a quantum channel.
Precisely, the basic concept of quantum information theory is that the
message is encoded in certain quantum states, which are transmitted through
some quantum channel, then the receiver decodes the quantum states at his
hand to retrieve the information. \ As a CPT, certainly the master equation
does reflect the time evolution of a density operator. Thereupon the master
equation defines a channel $\$$
\[
\$:\rho \rightarrow \$\left( \rho \right)
\]

Master equations intrinsically describe evolutions local in time, namely,
Markovian processes \cite{preskill}. Therefore, the quantum channels that
master equations define are quantum Markov channels.

According to Holevo-Schumacher-Westmoreland (HSW) theorem \cite{holevo2},
the one-shot classical capacity of the quantum channel $\$$\ is defined%
\begin{equation}
\chi \left( \$\right) \equiv \max_{\left\{ p_{i},\rho _{i}\right\} }\left[
S\left( \$\left( \sum_{i}p_{i}\rho _{i}\right) \right) -\sum_{i}p_{i}S\left(
\$\left( \rho _{i}\right) \right) \right]
\end{equation}%
where the maximum is over all ensembles $\left\{ p_{i},\rho _{i}\right\} $
of possible input states $\rho _{i}$ to the channel. Here $S\left( \rho
\right) =-Tr\left[ \rho \log \rho \right] $ is the von Neumann entropy. We
declare here that the basis of the logarithm function is 2 all through the
paper. In this case, $\chi \left( \$\right) $\ is in units of bit. The
procedure to calculate the channel capacity requires a maximization over all
the input states. So far as the continuous-variable quantum channel is
concerned, it is conjectured that a Gaussian mixture of coherent states,
namely, thermal state, achieves the Gaussian\ channel capacity $\chi \left(
\$\right) $ \cite{macchiavello}.

\subsection{The master equation and its solution}

Here we consider the case that monomodal electromagnetic field interacting
with the thermal reservoir, i.e., the damping harmonic oscillators coupled
to the squeezed thermal reservoir. The interaction Hamiltonian of the system
is \cite{scully}
\[
H=\hbar \sum_{k}g_{k}\left[ b_{k}^{+}ae^{-i(\nu -\nu
_{k})t}+a^{+}b_{k}e^{i(\nu -\nu _{k})t}\right]
\]%
where $a$ (and $a^{+}$) are the annihilation (and creation) operators of the
mode of interest. The operators $b_{k}^{+}$ and $b_{k}$ represent modes of
the reservoir that damp the field. When the modes $b_{k}$ are initially in a
squeezed vacuum, the evolution of the reduced density operator in the
interaction picture is described by the master equation given below \cite%
{scully,milburn}%
\begin{eqnarray}
\frac{d\rho }{dt} &=&\frac{\gamma }{2}(N+1)(2a\rho a^{+}-a^{+}a\rho -\rho
a^{+}a) \\
&&+\frac{\gamma }{2}N(2a^{+}\rho a-aa^{+}\rho -\rho aa^{+})  \nonumber \\
&&+\frac{\gamma }{2}M\left( 2a^{+}\rho a^{+}-a^{+}a^{+}\rho -\rho
a^{+}a^{+}\right)  \nonumber \\
&&+\frac{\gamma }{2}M^{\ast }\left( 2a\rho a-aa\rho -\rho aa\right)
\nonumber
\end{eqnarray}%
where $\gamma $ is the decaying rate, $N$ is the mean photon number of the
reservoir, and $M$ is the parameter somehow related to the squeezed vacuum
reservoir, respectivly.

The authors in \cite{lu} provide the explicit solution to the master
equation above. Assume the initial state is squeezed coherent state, namely,%
\[
\rho \left( 0\right) =S\left( \zeta \right) \left\vert \eta \right\rangle
\left\langle \eta \right\vert S^{+}\left( \zeta \right)
\]%
here $\left\vert \eta \right\rangle \left\langle \eta \right\vert $ is
coherent state, and $S\left( \zeta \right) $ is squeeze operator. Therefore
the final form of the output density operator is%
\[
\rho \left( t\right) =S\left( \zeta \right) \frac{\exp \left[ -\beta \left(
t\right) \left\vert \widetilde{\eta }\left( t\right) \right\vert ^{2}\right]
}{1+\beta \left( t\right) }\times \dsum\limits_{n=0}^{\infty }\left( \frac{%
\beta \left( t\right) }{1+\beta \left( t\right) }\right) ^{n}\frac{\left(
a^{+}\right) ^{n}}{n!}\left\vert \widetilde{\eta }\left( t\right)
\right\rangle \left\langle \widetilde{\eta }\left( t\right) \right\vert
a^{n}S^{+}\left( \zeta \right)
\]%
here%
\begin{equation}
\beta \left( t\right) =\frac{\beta }{\gamma }\left( 1-e^{-\gamma t}\right)
\end{equation}%
\begin{equation}
\widetilde{\eta }\left( t\right) =\frac{\eta e^{-\frac{\gamma }{2}t}}{%
1+\beta \left( t\right) }\equiv \eta f,f=\frac{e^{-\frac{\gamma }{2}t}}{%
1+\beta \left( t\right) }
\end{equation}%
where $\beta $ is a real number.

A unitary operator doesn't affect classical capacity of the quantum channel,
so we set $\zeta =0$. Hence the final solution has the form%
\begin{eqnarray*}
\rho \left( t\right) &=&\frac{\exp \left[ -\beta \left( t\right) \left\vert
\widetilde{\eta }\left( t\right) \right\vert ^{2}\right] }{1+\beta \left(
t\right) } \\
&&\times \dsum\limits_{n=0}^{\infty }\left( \frac{\beta \left( t\right) }{%
1+\beta \left( t\right) }\right) ^{n}\frac{\left( a^{+}\right) ^{n}}{n!}%
\left\vert \widetilde{\eta }\left( t\right) \right\rangle \left\langle
\widetilde{\eta }\left( t\right) \right\vert a^{n} \\
&=&\frac{1}{1+\beta \left( t\right) }e^{-\left( 1+\beta \left( t\right)
\right) f^{2}\left\vert \eta \right\vert ^{2}}e^{f\eta a^{+}}:e^{\frac{\beta
\left( t\right) }{1+\beta \left( t\right) }a^{+}a-a^{+}a}:e^{f^{\ast }\eta
^{\ast }a} \\
&=&\frac{1}{1+\beta \left( t\right) }e^{-\left( 1+\beta \left( t\right)
\right) f^{2}\left\vert \eta \right\vert ^{2}}:e^{-\frac{1}{1+\beta \left(
t\right) }a^{+}a+f\eta a^{+}+f^{\ast }\eta ^{\ast }a}:
\end{eqnarray*}

\subsection{The channel capacity}

Below we directly compute the channel capacity. Note that
\begin{eqnarray*}
\rho \left( t\right) &=&\frac{1}{1+\beta \left( t\right) }e^{-\left( 1+\beta
\left( t\right) \right) f^{2}\left\vert \eta \right\vert ^{2}}e^{f\eta
a^{+}}:e^{\frac{\beta \left( t\right) }{1+\beta \left( t\right) }%
a^{+}a-a^{+}a}:e^{f^{\ast }\eta ^{\ast }a} \\
&=&\frac{1}{\beta \left( t\right) }e^{-\left( 1+\beta \left( t\right)
\right) f^{2}\left\vert \eta \right\vert ^{2}}e^{f\eta a^{+}}:\int
d^{2}\alpha e^{-\frac{1+\beta \left( t\right) }{\beta \left( t\right) }%
\left\vert \alpha \right\vert ^{2}+\alpha a^{+}+\alpha ^{\ast
}a-a^{+}a}:e^{f^{\ast }\eta ^{\ast }a} \\
&=&\frac{1}{\beta \left( t\right) }e^{-\left( 1+\beta \left( t\right)
\right) f^{2}\left\vert \eta \right\vert ^{2}}e^{f\eta a^{+}}\int
d^{2}\alpha e^{-\frac{\left\vert \alpha \right\vert ^{2}}{\beta \left(
t\right) }}\left\vert \alpha \right\rangle \left\langle \alpha \right\vert
e^{f^{\ast }\eta ^{\ast }a} \\
&=&\frac{1}{\beta \left( t\right) }\int d^{2}\alpha e^{-\frac{\left\vert
\alpha -\beta \left( t\right) f\eta \right\vert ^{2}}{\beta \left( t\right) }%
}D\left( \alpha \right) \left\vert f\eta \right\rangle \left\langle f\eta
\right\vert D^{+}\left( \alpha \right)
\end{eqnarray*}

With this simplification, we can see $\rho \left( t\right) $\ is Gaussian.
Therefore we can safely arrive at the conclusion that the channel is a
Gaussian one.

Now we calculate the von Neumann entropy of $\rho \left( t\right) $. $\rho
\left( t\right) $ can be transformed into%
\begin{eqnarray*}
\rho \left( t\right) &=&\frac{1}{1+\beta \left( t\right) }e^{-\left( 1+\beta
\left( t\right) \right) f^{2}\left\vert \eta \right\vert ^{2}}:e^{-\frac{1}{%
1+\beta \left( t\right) }a^{+}a+f\eta a^{+}+f^{\ast }\eta ^{\ast }a}: \\
&=&\frac{1}{1+\beta \left( t\right) }:e^{-\frac{1}{1+\beta \left( t\right) }%
\left[ a^{+}-\left( 1+\beta \left( t\right) \right) f^{\ast }\eta ^{\ast }%
\right] \left[ a-\left( 1+\beta \left( t\right) \right) f\eta \right] }: \\
&=&\frac{1}{1+\beta \left( t\right) }D\left( -e^{-\frac{\gamma }{2}t}\eta
^{\ast }\right) :e^{-\frac{1}{1+\beta \left( t\right) }a^{+}a}:D^{+}\left(
-e^{-\frac{\gamma }{2}t}\eta ^{\ast }\right) \\
&=&\frac{1}{1+\beta \left( t\right) }D\left( -e^{-\frac{\gamma }{2}t}\eta
^{\ast }\right) e^{\ln \frac{\beta \left( t\right) }{1+\beta \left( t\right)
}a^{+}a}D^{+}\left( -e^{-\frac{\gamma }{2}t}\eta ^{\ast }\right) \\
&=&\frac{1}{1+\beta \left( t\right) }e^{\ln \frac{\beta \left( t\right) }{%
1+\beta \left( t\right) }\left( a^{+}-e^{\frac{\gamma }{2}t}\eta ^{\ast
}\right) \left( a-e^{\frac{\gamma }{2}t}\eta \right) }
\end{eqnarray*}

The von Neumann entropy of $\rho \left( t\right) $ is derived as%
\begin{eqnarray*}
S\left( \rho \left( t\right) \right) &=&-Tr\left( \rho \left( t\right) \log
\rho \left( t\right) \right) \\
&=&\left( 1+\beta \left( t\right) \right) \log \left( 1+\beta \left(
t\right) \right) -\beta (t)\log \beta (t)
\end{eqnarray*}

As previously stated, a Gaussian mixture of coherent states is assumed to
achieve the channel capacity \cite{macchiavello}. Therefore to attain the
channel capacity we have to compute the Gaussian mixture density operator of
$\rho \left( t\right) $, i.e. $\overline{\rho }\left( t\right) $ and it is
written as
\begin{equation}
\overline{\rho }\left( t\right) =\int d^{2}\eta p\left( \eta \right) \rho
\left( t\right)
\end{equation}%
where $p\left( \eta \right) =\frac{1}{\pi \overline{n}}e^{-\frac{\left\vert
\eta \right\vert ^{2}}{\overline{n}}}$, with $\overline{n}$ being the mean
photon number at the input of the channel. Substitute $\rho \left( t\right) $
into the equation above, it is easy to obtain%
\[
\overline{\rho }\left( t\right) =\int \frac{d^{2}\eta }{\pi \overline{n}}e^{-%
\frac{\left\vert \eta \right\vert ^{2}}{\overline{n}}}e^{-\left( 1+\beta
\left( t\right) \right) \left\vert f\right\vert ^{2}\left\vert \eta
\right\vert ^{2}}:e^{-\frac{1}{1+\beta \left( t\right) }a^{+}a+f\eta
a^{+}+f^{\ast }\eta ^{\ast }a}:
\]

Straightforward calculation shows that the integral gives rise to%
\[
\overline{\rho }\left( t\right) =\frac{1}{1+\beta \left( t\right) +\overline{%
n}e^{-\gamma t}}:e^{-\frac{1}{1+\beta \left( t\right) +\overline{n}%
e^{-\gamma t}}a^{+}a}:
\]

Naturally the von Neumann entropy of $\overline{\rho }\left( t\right) $ is%
\begin{eqnarray*}
S\left( \overline{\rho }\left( t\right) \right) &=&-Tr\left( \overline{\rho }%
\left( t\right) \log \overline{\rho }\left( t\right) \right) \\
&=&-\log \frac{1}{1+\beta \left( t\right) +\overline{n}e^{-\gamma t}}-\left(
\beta \left( t\right) +\overline{n}e^{-\gamma t}\right) \log \frac{\beta
\left( t\right) +\overline{n}e^{-\gamma t}}{1+\beta \left( t\right) +%
\overline{n}e^{-\gamma t}} \\
&=&\left( 1+\beta \left( t\right) +\overline{n}e^{-\gamma t}\right) \log
\left( 1+\beta \left( t\right) +\overline{n}e^{-\gamma t}\right) -\left(
\beta \left( t\right) +\overline{n}e^{-\gamma t}\right) \log \left( \beta
\left( t\right) +\overline{n}e^{-\gamma t}\right)
\end{eqnarray*}

According to definition (1), the channel capacity
\begin{eqnarray}
\chi \left( \$\right) &\equiv &\max_{\left\{ p_{i},\rho _{i}\right\} }\left[
S\left( \$\left( \sum_{i}p_{i}\rho _{i}\right) \right) -\sum_{i}p_{i}S\left(
\$\left( \rho _{i}\right) \right) \right] \\
&=&S\left( \overline{\rho }\left( t\right) \right) -\int d^{2}\eta p\left(
\eta \right) S\left( \rho \left( t\right) \right)  \nonumber \\
&=&\left[ \left( 1+\beta \left( t\right) +\overline{n}e^{-\gamma t}\right)
\log \left( 1+\beta \left( t\right) +\overline{n}e^{-\gamma t}\right)
-\left( \beta \left( t\right) +\overline{n}e^{-\gamma t}\right) \log \left(
\beta \left( t\right) +\overline{n}e^{-\gamma t}\right) \right]  \nonumber \\
&&-\left( 1+\beta \left( t\right) \right) \log \left( 1+\beta \left(
t\right) \right) +\beta (t)\log \beta (t)  \nonumber
\end{eqnarray}

\section{Discussions}

According to the analysis in Sec II, the capacity of the noisy Markov
quantum channel is
\begin{eqnarray}
\chi \left( \$\right) &=&\left[ \left( 1+\beta \left( t\right) +\overline{n}%
e^{-\gamma t}\right) \log \left( 1+\beta \left( t\right) +\overline{n}%
e^{-\gamma t}\right) -\left( \beta \left( t\right) +\overline{n}e^{-\gamma
t}\right) \log \left( \beta \left( t\right) +\overline{n}e^{-\gamma
t}\right) \right] \\
&&-\left( 1+\beta \left( t\right) \right) \log \left( 1+\beta \left(
t\right) \right) +\beta (t)\log \beta (t)  \nonumber
\end{eqnarray}%
where $\overline{n}$ is the mean photon number at the input of the channel,
the classical equivalent counterpart of which is the energy constraint of
the information transmission. $\gamma $ is the decaying rate, and $\beta
\left( t\right) $ is the fluctuations of the thermal reservoir.

Substitute Eq.(3) into the equation above, it is easy to obtain%
\begin{eqnarray*}
\chi \left( \$\right) &=&\left( 1+\frac{\beta }{\gamma }\left( 1-e^{-\gamma
t}\right) +\overline{n}e^{-\gamma t}\right) \log \left( 1+\frac{\beta }{%
\gamma }\left( 1-e^{-\gamma t}\right) +\overline{n}e^{-\gamma t}\right) \\
&&-\left( \frac{\beta }{\gamma }\left( 1-e^{-\gamma t}\right) +\overline{n}%
e^{-\gamma t}\right) \log \left( \frac{\beta }{\gamma }\left( 1-e^{-\gamma
t}\right) +\overline{n}e^{-\gamma t}\right) \\
&&-(1+\frac{\beta }{\gamma }\left( 1-e^{-\gamma t}\right) )\log (1+\frac{%
\beta }{\gamma }\left( 1-e^{-\gamma t}\right) ) \\
&&+\left( \frac{\beta }{\gamma }\left( 1-e^{-\gamma t}\right) \right) \log
\left( \frac{\beta }{\gamma }\left( 1-e^{-\gamma t}\right) \right)
\end{eqnarray*}

Here $\chi \left( \$\right) $ is in units of bit; $\overline{n}$ is a
dimensionless value; $\gamma $ and $\beta $ are in units of $s^{-1}$; $t$ is
always in units of $s$.

If $\overline{n}$ ranges from $1$ to $10$ while $%
\gamma $ and $\beta $ are fixed as $0.1$ and $0.01$, respectively, we have
the following diagram figure 1.;

\begin{figure}[h]
\begin{center}
\includegraphics[width=0.4\textwidth]{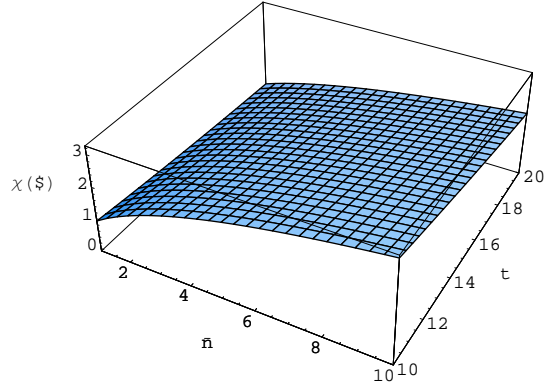}
\end{center}
\caption[bf]{$\overline{n}$ ranges from $1$ to $10$ while $\protect\gamma $
and $\protect\beta $ are fixed as $0.1$ and $0.01$, respectively.}
\end{figure}

Assume $\overline{n}$ is fixed to $5$ and $\gamma $ fixed to $0.1$, while $%
\beta $ runs from $0.01$ to $0.1$, then we have the following diagram figure
2.;

\begin{figure}[h]
\begin{center}
\includegraphics[width=0.4\textwidth]{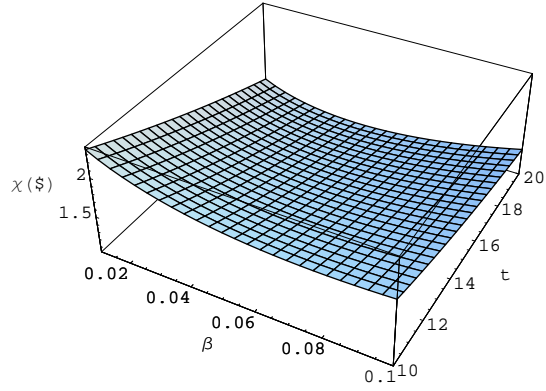}
\end{center}
\caption[bf]{$\overline{n}$ is fixed to $5$ and $\protect\gamma $ fixed to $%
0.1$, while $\protect\beta $ runs from $0.01$ to $0.1$.}
\end{figure}

If $\overline{n}$ is fixed to $5$ and $\beta $ is fixed to $0.01$, while $%
\gamma $ ranges from $0.1$ to $0.5$, then the diagram figure 3. below is
obtained.

\begin{figure}[h]
\begin{center}
\includegraphics[width=0.4\textwidth]{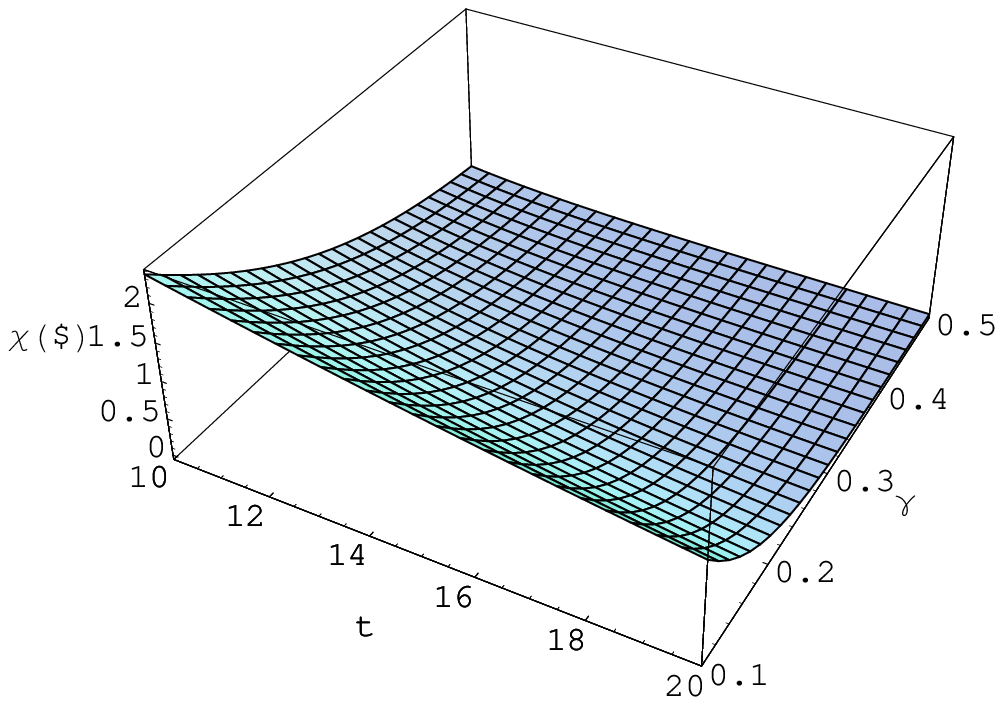}
\end{center}
\caption[bf]{$\overline{n}$ is fixed to $5$ and $\protect\beta $ is fixed to
$0.01$, while $\protect\gamma $ ranges from $0.1$ to $0.5$.}
\end{figure}

As these diagrams indicate, the channel capacity is proportional to the mean
photon number at the input of the channel, yet decreases with the increase
of time and intensity of noise $\beta \left( t\right) $. Here $\overline{n}$
is considered as the signal to the channel. The larger the signal, the
larger the channel capacity. That the Markov quantum channel is noisy lies
in the fact that $\overline{n}$ is decaying with the increase of the time.
Our results are quite rational.

\section{Fidelity vs. channel capacity}

In quantum information theory, the fidelity is of great significance, which
discusses the similarity between the input density operator and the output
density operator. In some aspect, it evaluates how successful the message is
transmitted and how well the channel preserves the information \cite%
{peters,j,lidar,oh}. The fidelity $\digamma $\ of the channel is calculated
as%
\begin{eqnarray*}
\digamma \left( \eta \right) &=&\left\langle \eta \right\vert \rho \left(
t\right) \left\vert \eta \right\rangle \\
&=&\left\langle \eta \right\vert \frac{1}{1+\beta \left( t\right) }:e^{-%
\frac{1}{1+\beta \left( t\right) }\left[ a^{+}-\left( 1+\beta \left(
t\right) \right) f^{\ast }\eta ^{\ast }\right] \left[ a-\left( 1+\beta
\left( t\right) \right) f\eta \right] }:\left\vert \eta \right\rangle \\
&=&\frac{1}{1+\beta \left( t\right) }e^{-\left( 1+\beta \left( t\right)
\right) f^{2}\left\vert \eta \right\vert ^{2}+f\left\vert \eta \right\vert
^{2}+f^{\ast }\left\vert \eta \right\vert ^{2}-\frac{1}{1+\beta \left(
t\right) }\left\vert \eta \right\vert ^{2}} \\
&=&\frac{1}{1+\beta \left( t\right) }\exp \left[ -\frac{\left( e^{-\frac{%
\gamma }{2}t}-1\right) ^{2}}{1+\beta \left( t\right) }\left\vert \eta
\right\vert ^{2}\right]
\end{eqnarray*}

And the average value of $\digamma $, i.e., $\overline{\digamma }$ is%
\begin{eqnarray*}
\overline{\digamma } &=&\int \frac{d^{2}\eta }{\pi \overline{n}}e^{-\frac{%
\left\vert \eta \right\vert ^{2}}{\overline{n}}}\frac{1}{1+\beta \left(
t\right) }\exp \left[ -\frac{\left( e^{-\frac{\gamma }{2}t}-1\right) ^{2}}{%
1+\beta \left( t\right) }\left\vert \eta \right\vert ^{2}\right] \\
&=&\frac{1}{1+\beta \left( t\right) +\overline{n}\left( e^{-\frac{\gamma }{2}%
t}-1\right) ^{2}}
\end{eqnarray*}

It is clear that the average fidelity decreases with growing $\overline{n}$,
nevertheless, the capacity increases with growing $\overline{n}$.
Mathematically, we hold that there exists particular $\overline{n}_{opt}$ at
which the average fidelity and the capacity are optimal, namely, certain
channel capacity can be achieved with reasonably high fidelity. Below we
derive the criterion for this situation.

Define $\Theta $ as%
\[
\Theta =\overline{\digamma }\times \chi \left( \$\right)
\]%
then $\overline{n}_{opt}$ must saturate the following equation
\[
\frac{\partial \Theta }{\partial \overline{n}}=0
\]%
that is to say, $\overline{n}_{opt}$ satisfies the equation%
\begin{eqnarray}
&&a(1+\beta (t))\log (1+\beta (t))-a\beta (t)\log \beta (t) \\
&=&(a\beta (t)-(1+\beta (t)))\log b-(a-1)(1+\beta (t))\log (1+b)  \nonumber
\end{eqnarray}%
where%
\[
a=\left( e^{\frac{\gamma }{2}t}-1\right) ^{2}
\]%
\[
b=\beta (t)+\overline{n}e^{-\gamma t}
\]

It is evident that any signal input $\overline{n}$ saturates the criterion
equation (8) above gives rise to mutually optimal channel capacity and
transmission fidelity.

\section{Conclusions}

In this paper we adopt a method to compute the quantum Markov channel of
continuous variables. We use this method to calculate explicitly a noisy
quantum channel with a Gaussian-distributed noise. Physically, the process
is a monomodal electromagnetic field decaying inside a cavity. From the
quantum information's point of view, it is a field propagating at the
prensence of reservoir.

We give the explicit form of the capacity of the channel. The channel
capacity is proportional to the input signal $\overline{n}$, while decaying
with the increase of time t and decaying rate $\gamma $ and the noise $\beta
$.

On the other hand, the fidelity of the channel, which is a tag of the
success of the information transmission, decreases with larger $\overline{n}$%
. Hence, there is tradeoff between fidelity and capacity. We derive the
criterion when these two variables can achieve balance.

Quantum channels of continuous variables have been an important issue. We
hope our research can shed light on this subject.

\end{document}